\documentstyle[11pt,leqno]{article}

 \newcommand{\beq}{\begin{equation}}                       
 \newcommand{\eeq}{\end{equation}}                         
 \newcounter{nt}[section]                                  
 \newcounter{nl}[section]                                  
 \date{}                                                   
 \textwidth  
12.5cm

\begin{document}

\vspace*{14mm}
\begin{center}
{ \Large \bf Non linear problems in dissipative models }
\end{center}

\vspace{2mm}
\begin {center}
{\normalsize Nota di M.De Angelis, G. Fiore. P.Renno}
   \footnote{ Facolt\`{a} di
             Ingegneria, Dip. Mat. Appl. "R. Caccioppoli",
 via Claudio 21, 80125, Napoli. E-mail: modeange@unina.it; gaetano.fiore@unina.it; renno@unina.it}
\end{center}

\begin {center}
{\small Presentata dal Socio  Pasquale Renno

(Adunanza del 10 Giugno 2005)}

\end{center}

\vspace{3mm}\noindent
{\small \it Key words:} {\small  Viscoelastic models, Superconductivity, Boundary-layer, Partial dif  \hspace*{20mm}ferential equations }

\vspace{8mm}\noindent
{ \small \bf Abstract-}{\small \@ Aim of the paper is the qualitative analysis of a quasi-linear parabolic third order equation, which describes the evolution in a large class of dissipative models. As examples of some  typical boundary problems, both Dirichlet's  and Neumann's type boundary conditions are examined. In the linear case, the related Green functions are explicitly determined, together with  rigorous estimates of their behavior when the parameter of dissipation  $ \varepsilon $ is vanishing. These results are basic to study the integral equations to which the non linear problems can be reduced. Moreover, boundary layer estimates  can be determined too.}

\vspace{8mm}\noindent
{ \small \bf Riassunto-}{\small  \@ Oggetto del lavoro e' l'analisi qualitativa di un'equazione parabolica quasi lineare del terzo ordine, che descrive l'interazione tra propagazione ondosa e diffusione in ampie classi di modelli dissipativi. Quali esempi di problemi al contorno, vengono prese in esame condizioni sia di tipo Dirichlet che di tipo Neumann. Nel caso lineare, si determinano esplicitamente le relative funzioni di Green,  insieme ad alcune stime rigorose  del loro comportamento quando il parametro $\varepsilon$ della dissipazione tende a zero. Queste stime costituiscono la premessa per l'analisi delle equazioni integrali cui e' possibile ridurre i problemi al contorno non lineari.}

\vspace{4mm}
 \section{  Introduction}
 \setcounter{equation}{0}

\hspace{5.1 mm}

A great deal of models of applied sciences are described by the parabolic equation:

\vspace{3mm}

\begin {equation}    \label{11}
 {\cal{L}}_\varepsilon  u = \varepsilon u_{xxt}
+ c^2 u_{xx}  - u_{tt}  - 2 a u_t  = -f. 
\end{equation}

\vspace{3mm}

 The constants 
$ a, c^2, \varepsilon $ are all positive and they assume various meanings according to physical problems. As for the source $f$, it can be linear or not.

For instance, the equation (\ref {11}) is involved in the generalized Maxwell- Cattaneo system of equations \cite{jp}-\cite{ps},  in problems of viscoelastic media of Kelvin-Voigt type \cite {r},  or for the study of solids at very low temperatures \cite {jcl}.  
Further applications arise in the study of viscoelastic plates with memory, when the relaxation function is given by  exponential functions. (\cite {cdf} and  references therein).

 A typical example of the non linear case   
is the {\it
perturbed sine-Gordon equation}  which models the flux dynamics in Josephson junctions in 
 superconductivity  \cite{scott}\cite{bp}.  In this case, the terms $\varepsilon u_{xxt}$ and $a u_t$ characterize the
 dissipative normal electron current flow along and across the junction.
 
As for the practical applications of superconductors, many areas are involved. In  medicine, for instance,  Magnetic  Resonance Imaging (MRI) has been used since  1977 and    is still  improving \cite{ebphkba}.  Referring to the  electric power systems, the high temperature superconductor cables are likely to lead to a lot of benefits as regards the  current carrying capacity and for reducing electrical losses.\cite{thjlmoso}\cite{n}.

As for typical  boundary value problems related to the equation (\ref{11}), both the Dirichlet conditions and Neumann conditions have interest for  practical applications. For instance, in superconductivity,  the first case  can be referred to  periodic conditions according to  annular geometry of junction \cite {parm}{\cite {cln}, while in the other case, the phase gradient, proportional to the magnetic field, is specified.(\cite{scr}-\cite{fppcss}). When the source term $f$ is {\it linear}, all these problems can be explicitly solved by means of the Fourier method. The solutions are determined in  sect.2-3, together with the related Green functions $ G_\varepsilon , K_\varepsilon $.

When the function $f$ is  {\it non linear}, then $ G_\varepsilon , K_\varepsilon $ represent the kernels of the integral equations to which the above mentioned boundary value problems can be reduced. For this, a rigorous analysis of the behavior of these kernels when $\, \varepsilon \rightarrow 0 $ and $\, t \rightarrow \infty\,$ is achieved in sect 4.  At last, as first application, the influence of the dissipation on the wave behavior is estimated by an asymptotic approximation uniformly valid  also for large  $\,t\,$ (sect.5).


\vspace{5.1mm}

\section {Statement of the problem}
 \setcounter{equation}{0}

\hspace{5.1mm}

\vspace{3mm}
     If  $ u _\varepsilon (x,t)$  is a function defined in the strip

\begin {center}
$  \Omega = \{(x,t) : 0 \leq x
\leq
\pi, \  \ t \,\geq \,0 \}$,
\noindent

\end {center}
 let ${\cal P}_\varepsilon  $  the  initial- boundary value
problem related to equation (\ref{11}) with conditions

\vspace{3mm}
\begin{equation}                               \label{21}
  \begin{array}{lll}
   u_\varepsilon (x,0)=f_0(x), \ \  \partial_t u_\varepsilon
   (x,0)=f_1(x),
& x\in [0,\pi],\vspace{2mm}  \\
\end{array}
\end {equation}   
  
\begin {equation}                   \label{22}
\begin{array}{ll}
u_ \varepsilon(0,t)=\varphi(t), \  \  \ \ \ \ \ u_\varepsilon
    (\pi,t)=\psi(t), & t \geq 0,
   \end{array}
 \end{equation}

\vspace{3mm}
\noindent
where $ f_0, f_1, \psi, \varphi $ are arbitrary date.

\vspace{3mm}
The boundary conditions (\ref{22}) represent only an example of the analysis we are going to apply. Equally, {\em flux-boundary } conditions or {\em mixed-boundary } conditions can be considered too. So, another example is given by the problem ${\cal H}_\varepsilon  $ defined in $\Omega  $ by  (\ref{11})-(\ref{21}) together with the Neumann conditions

\vspace{3mm}

\begin {equation}                   \label{23}
\begin{array}{ll}
\partial _x \,u_ \varepsilon(0,t)=\varphi_1(t), \  \  \ \ \ \ \ \partial _x \,u_\varepsilon
    (\pi,t)=\psi_1(t), & t\geq 0.
   \end{array}
 \end{equation}

\vspace{3mm}

When $\varepsilon \equiv 0$, the parabolic equation (\ref{11}) turns into the hyperbolic telegraph equation

 \vspace{3mm}
\begin{equation}                               \label{24} 
  {\cal
L}_0  u_0 \equiv ( c^2 \, \partial_{xx} - \partial_{tt} -2\, a  \,\partial_t ) u_0 = - \bar f(x,t,u_0)
\eeq

\vspace{3mm}\noindent 
and the problem   ${\cal
P}_\varepsilon $ changes into a problem   ${\cal
P}_0 $ for $ u_0(x,t)$ which has the same initial-boundary conditions (\ref{21}) - (\ref{22}) of ${\cal
P}_\varepsilon $. When the source term $\bar f$ of (\ref{24}) is {\em linear} $(\bar f= \bar f (x,t))$, ${\cal
P}_0 $ is explicitly solved by means of the well-known Green function:

\vspace{3mm}

\begin{equation}                \label {25}
G_0 (x,\xi,t)\,\,= \,\, \frac{2}{\pi } \,\,  \,e^{-a \,t } \,\ \sum
_{n=1}^{\infty} \,\,  \,\, \frac{ \sin \,(t \,\, \sqrt{c^2 n^2 - a^2}\, \,)}{    \sqrt{c^2 n^2 - a^2}}   
\, \, \sin  (nx)  \,\, \sin
(n\xi)
\end{equation}

\vspace{3mm} 

In order to estimate the influence of the dissipative term $\, \varepsilon \, u_{xxt}\,$ on the wave behavior of $\,u_0\,$, the difference

\vspace{3mm}

\begin{equation}                \label {26}
 v(x,t) = u_\varepsilon - u_0,
\end{equation}

\vspace{3mm}\noindent
is to be evaluated and so the following {\em problem   $
\Delta $ }  must be analyzed

\vspace{3mm}
  \beq                                                     \label{27}
  \left \{
   \begin{array}{lll}
    \varepsilon \, v_{xxt}+c^2\, v_{xx} -v_{tt}-2av_t  = - F(x,t, u_0 ,v )   & (x,t) \in \Omega \\
\\
 v(x,0)=0, \  \   v_t(x,0)=0, \  & x \in [0,\pi], \vspace{2mm}  \\
\\
    v(0,t)=0, \  \ v(\pi,t)=0,  & t\geq 0.
   \end{array}
  \right.
 \eeq 

\vspace{3mm}
\noindent
The source term $F$ is given by

\vspace{3mm}
\beq                                       \label{28}
 F= f(x,t,u_o+v) -\bar f(x,t,u_0) + \varepsilon \,{u_{0}}_{xxt},
\eeq

\vspace{3mm}\noindent
while, in the linear case, it is $\,f=\bar f\,$ and $ F= \varepsilon \,{u_{0}}_{xxt}$.

\vspace{3mm}
As for the problem ${\cal H}_\varepsilon $, instead of $(\ref{27})_3 $ the following conditions

\vspace{3mm}
  \beq                                                     \label{29}
    v_x(0,t)=0, \  \ v_x(\pi,t)=0,  \,\,\,\,\,\,\,t\geq 0,
    \eeq

\vspace{3mm}\noindent
must be specified.

\vspace{5.1mm}

\section {Linear case and explicit solutions}
 \setcounter{equation}{0}

\hspace{5.1mm}

\vspace{3mm}

\vspace{3mm} Let $ \hat z(s)$ the Laplace-trasform  of the function $z(t) $ and let 

\vspace{3mm}  

\begin{equation}             \label {31}
\sigma (s) = \sqrt {\frac{s^2 +2as}{\varepsilon s +c^2}}.
\end{equation}

\vspace{3mm}\noindent
When $F $ is linear and the Laplace trasform is applied to the problem $\Delta$, the transform $\hat v(x,s)$ of the solution $v(x,t)$ is given by

\vspace{3mm}

\begin{equation}                  \label {32}
\hat v(x,s) = \int_{0}^{\pi}
 \hat{G}_\varepsilon
(x,\xi,s) \ \hat F(\xi,s) \, d\xi,
\end{equation}

\vspace{3mm}\noindent where

\vspace{3mm}
\begin{equation}                         \label{33}
\hat{G}_\varepsilon =  \, \frac{1}{2(\varepsilon s +c^2)} \,\,\, [\,\hat{g} (|x-\xi|, \sigma) -
\hat{g} (|x+\xi|, \sigma)\,]
\end{equation}

\vspace{3mm}\noindent and 

\vspace{3mm}

\begin{equation}             \label {34}
\hat{g}(y,\sigma) = \frac{cosh\,[(\pi-y)\,\sigma]}{
\sigma  \, senh (\pi\sigma)}.
\end{equation}

\vspace{3mm}
\noindent
But, for $y \in [0, 2 \pi]$, it results \cite{g}:

\vspace{3mm}
\beq                             \label{35}
\hat g(y,\sigma) \,= \frac{1}{\pi \sigma^2} \,+ \frac{2}{\pi} \, \, \ \,\sum _{n=1}^{\infty} \,\, \frac{\,\cos\,(ny)}{n^2 + \sigma ^2}
\eeq
 
\vspace{3mm}
\noindent
and $ \cos\,(n|x-\xi|) \,\, - \,\, \cos\,(n|x+\xi|) \, = 2 \,\sin \,(nx)\,\, \sin\,(n \xi).$ So, by (\ref{33})- (\ref{35}), it follows

\begin{equation}                         \label{36}
\hat{G}_\varepsilon(x,\xi, s) =  \frac{2}{\pi} \, \, \ \,\sum _{n=1}^{\infty} \,\, \, \frac{\,\sin\,(n\xi) \,\, \sin \, (nx)}{s^2 + 2as + (\varepsilon s+c^2 ) n ^2},
\end{equation}

\vspace{3mm}\noindent
which represents the  ${\cal L}$ - transform of the Green function related to problem $\Delta$.

By means of elementary formulae, one deduces 

\vspace{3mm}
\begin{equation}                \label {37}
G_\varepsilon (x,\xi, t) = \frac{2}{\pi } \,\,  \sum
_{n=1}^{\infty} \, \,
 h_n\,(t, \varepsilon ) \,\, \sin  (nx)  \,\, \sin
(n\xi),
\end{equation}

 \vspace{3mm}\noindent with 
\vspace{3mm}
\beq                           \label{38}
h_n(t,\varepsilon)\,\,=e^{-(\frac{\varepsilon}{2}n^2 +a)\, t } \ \   \frac{\sin \,[\,
 t \,\sqrt {c^2n^2- {(\frac {\varepsilon} {2}n^2 +a)^2}}\,]
}{ \sqrt {c^2 n^2- (\frac {\varepsilon }{2}n^2 +a)^2}}.
\eeq

\vspace{3mm}
Then, the esplicit solution $v$ of the problem $\Delta$ is:

\vspace{3mm}
\begin{equation}                  \label {39}
v (x,t)=   \int_{0}^{t} \, \,d \tau \,\,\int_{0}^{\pi} F (\xi,\tau)\,\, \ {G}_\varepsilon (x,\xi, t-\tau) \,d\xi
\eeq

\vspace{3mm}
\noindent 
with the Green function $G_\varepsilon$ defined by (\ref{37}),(\ref{38}).

The formal analysis developed so far can be justified as follows. Referring to (\ref{39})- (\ref{37}), the terms

\vspace{3mm}
\begin{equation}                  \label {310}
F_n\, (t)\, = \, \frac{2}{\pi} \, \,   \int_{0}^{\pi} F (\xi,t)\,\, \sin \,(n\xi) \, \, d\xi
\eeq

\vspace{3mm}
\noindent 
represent the Fourier coefficients of the sine series of the function $F(x,t)$:

\vspace{3mm}
\beq                          \label{311}
F(x,t) =  \sum
_{n=1}^{\infty} \, \,
 F_n\,(t) \,\, \sin  (nx) \
\end{equation}

\vspace{3mm}\noindent
and the rapidity of pointwise convergence of this series depends, of course,  on the properties of the source $F$. For instance, it can be sufficiently assumed that $\, F, \,\,F_x,\,\, F_{xx}\,$ are continuous in $(0,\pi)$ and more

\vspace {3mm}
  \beq                               \label{312}
 F(0,t)=F(\pi,t)=0.
 \eeq

\vspace{3mm}
Then, the convergence of (\ref{311}) is uniform everywhere in $[0,\pi]$ and, further, it results:

\vspace{3mm} 

\begin{equation}                  \label {313}
F_n\, (t)\, = \, - \, \frac{1}{n^2} \,\, \frac{2}{\pi} \, \,   \int_{0}^{\pi} F_{\xi\xi} (\xi,t)\,\, \sin \,(n\xi) \, \, d\xi.
\eeq

\vspace{3mm}
As consequence, if one puts:

\vspace{3mm}
\begin{equation}                  \label {314}
v_n(t) \, = \,\,  h_n * \, F_n = \,\, \int_{0}^{t} F_n (\tau)\,\,\, h_n(t-\tau) \, \, d\tau,
\eeq

\vspace{3mm}
\noindent  
the solution (\ref{39}),(\ref{37}) represents the Fourier sine expansion
of $ v(x,t)$:

\vspace{3mm}
\beq                          \label{315}
v(x,t) =  \sum
_{n=1}^{\infty} \, \,
 v_n\,(t ) \,\, \sin  (nx).
\end{equation}

\vspace{3mm}

\vspace{3mm}
{\bf Theorem 3.1}-
{\em When }$F(x, \cdot) \in C^2\, ( \Lambda)$ {\em  and satisfies }\ref{312}, {\em the solution } $v(x,t) $ {\em of the problem }$\Delta$ {\em can be given the form }:

\vspace{3mm}
\begin{equation}                  \label {316}
v (x,t) \, = \, - \,   \int_{0}^{t} \, \,d \tau \int_{0}^{\pi} F_{\xi\xi} \, (\xi,\tau)\,\, \ {H}_\varepsilon \, (x,\,\xi,\, t-\tau) \,\, d\xi,
\eeq

\vspace{3mm}
\noindent
{\em where} $H_\varepsilon$ {\em is}

\vspace{3mm}
\begin{equation}                \label {317}
H_\varepsilon (\xi, x, t) = \frac{2}{\pi } \,\,  \sum
_{n=1}^{\infty} \, \, \frac{
 h_n\,(t, \varepsilon )}{n^2} \,\, \sin  (nx)  \,\, \sin
(n\xi)
\end{equation}

 \vspace{3mm}\noindent
 {\em and the convergence of the series is uniform everywhere in} $[0,\pi]$. 
\hbox{}\hspace*{3mm}
\rule{1.85mm}{1.85mm}

\vspace{3mm}
{\bf Remark 3.1}- Theorem 3.1 can be applied also to the problem $ {\cal H}_\varepsilon$, provided that the Green function $G_\varepsilon$ is substituted by the following function:

\vspace{3mm}
\begin{equation}                \label {318}
K_\varepsilon (\xi, x, t) =\, \,\frac{h_0(t)}{\pi}\,+\,\frac{2}{\pi } \,\,  \sum
_{n=1}^{\infty} \, \, 
 h_n\,(t, \varepsilon ) \,\, \cos  (nx)  \,\, \cos
(n\xi),
\end{equation}


 \vspace{5.1mm}

\section {Estimates and properties of the series $H_\varepsilon$}
 \setcounter{equation}{0}

\hspace{5.1mm}

\vspace{3mm}

The arguments of sine functions in (\ref{38})\@ are real parameters when \@$ K_1 \leq n\, \leq\, K_2$, with

\vspace{3mm}
\beq                  \label{41}
  K_1 = \,\, \frac{c}{\varepsilon}\,\,(\, 1- \sqrt{ 1- \frac{2 \, a \, \varepsilon }{c^2}}\,\,\,\,), \  \  \   K_2 = \,\, \frac{c}{\varepsilon}\,\,(1+ \sqrt{ \,1- \frac{2 \, a \, \varepsilon }{c^2}}\,\,\,\,).
\eeq

\vspace{4mm}
So, if   $a\,<\,c$   \@ and  $  N \,\equiv \,[K_2]$ , \@ the $h_n$ 's in (\ref{38}) contain trigonometric functions for $1 \, \leq \, n\, \leq  \, N $ and hyperbolic functions for $ n \, \geq \, N+1$. Otherwise, if $a>c$, the trigonometric case is related only to  $[K_1] \, \leq \, n\, \leq  \, N $. This distinction is unimportant to what we are going to demonstrate; however it holds also for the Green function $G_0$ defined by (\ref{25}) and related to the problem ${\cal P}_0 $.

\vspace{3mm}

 Let $ \, g_n(x,\xi) = (2/\pi) \,\, \sin(nx) \,\, \sin (n\xi)\,\,$ and consider the series 

\vspace{3mm}

\vspace{3mm}

\begin{equation}                \label {42}
H_0 \,\,= \,\,\,e^{-a \,t } \,\   \sum
_{n=1}^{\infty} \,\,  \,\, \frac{\sin \,(\, t\,\sqrt{c^2 n^2 - a^2}\,)\,}{ \,\,n^2\,\,   \sqrt{c^2 n^2 - a^2}}   
\, \,g_n(x,\xi)
\end{equation}

 \vspace{3mm}\noindent
 deduced from $H_\varepsilon$ setting formally $\varepsilon \equiv 0$.

In order to estimate the difference $\,H_\varepsilon -\,H_0$, let  

\vspace{3mm}
\begin{equation}                \label {43}
A_n \,\,= a+ \frac{\varepsilon}{2} \,\, n^2 ; \,\, \,\, B_n^2 \,\, = {c^2 n^2 - A_n^2};\,\,\,\,\, b_n^2 =\,\,{c^2 n^2 - a^2}   
\end{equation} 

\vspace{3mm}\noindent
and

\begin{equation}                \label {44}
 r_n =\, \frac{h_n (t,\varepsilon)}{n^2}\,-\, \frac{h_n(t,0)}{n^2}=  e^{-A_n \,\,t} \, \frac{sin (B_n \,t)}{ n^2 \,B_n} \,-\, e^{-at} \,  \frac{sin (b_n \,t)}{ n^2 \,b_n}  
\end{equation}

 \vspace{3mm}\noindent
with $h_n(t,\varepsilon)$ defined by (\ref{38}). It results:

\vspace{3mm}
\beq            \label{45}
 H_\varepsilon -H_0 \,= R_1 \, + R_2 \, =   \sum ^N _{n=1} \,\, \,  r_n ( t,\varepsilon ) \,\, g_n \,+\,\sum ^\infty _{n=N+1} \,\, \,  r_n ( t,\varepsilon ) \,\, g_n.  
\eeq

\vspace{3mm}

 If \@ $c_0$ \@ denotes the Euler constant \@$( c_0 \, \simeq \,0,5773)$\@   
  and  $k$ an {\em arbitrary} constant such that \@ $ 0\,<\,k\,<\,1\,$,  let  $\,b^2_1=c^2-a^2$ and

\vspace{3mm}             
\beq                       \label{46}
\rho (t) \, = \, \frac{t}{b_1} \, (2+at), \ \ \ c_1(\varepsilon) =\frac{\varepsilon + 2 c_0 \, c + (2c)^{2-k} \,\, \varepsilon ^{k-1}}{\pi \, c \, (1-k)}.
\eeq

\vspace{3mm}\noindent 
Further, let $c_2 \equiv \, (1/3)\,\,max (1,\pi /cb_1).$ Then one has:

\vspace{3mm}
 {\bf Lemma 4.1} - {\em For all } $ t\,\geq 0, \,\,x \in [0,\pi]\,$,{\em  \@ when }$\varepsilon$ {\em is vanishing, the following estimates hold:}

\vspace{3mm}
\beq                             \label{47}
 \mid R_1 \mid \,\, \leq \,\, \varepsilon \,\, c_1 (\varepsilon) \,\, \rho ( t) \, e^{-at} 
\eeq

 \vspace{3mm}

\beq                             \label{48}
 \mid R_2 \mid \,\, \leq \,\, \varepsilon \,\, c_2  \,\, [  e^{-at} \,+ \, \theta \,e^{-\frac{c^2}{2}\,\theta}], 
\eeq

 \vspace{3mm}\noindent
{\em where} $\theta $ {\em denotes the fast time} $ t/\varepsilon$ {\em and }$ \varepsilon \, c_1 (\varepsilon)$ {\em vanishes with arbitrary order } $ k <1$.

\vspace{3mm}
{\bf Proof}- Referring to the trigonometric terms related to $R_1$, defined in (\ref{45}), by means of the Laplace transform, by (\ref{44}) one deduces that

\vspace{3mm}
\begin{equation}                \label {49}
\hat r_n(s, \varepsilon)= \, -\,\frac{\varepsilon  }{b_n} \,\, \frac{b_n}{(s+a)^2 \,+\,b_n^2}\,\,\,\frac{s}{(s+A_n)^2\,+\,B_n^2} 
\end{equation}

 \vspace{3mm}\noindent 
hence

 \vspace{3mm}
\begin{equation}                \label {410}
 r_n( t,\varepsilon) =  \frac{\varepsilon  }{b_n}  [ e^{-at} \sin (b_n t)] * [ e^{-A_n t}  ( \frac {A_n}{B_n}  \sin (B_n \,t)- \cos (B_n  t))]. 
\end{equation}

\vspace{3mm}\noindent
When this convolution is made explicit, by elementary estimates one has

 \vspace{3mm}
\begin{equation}                \label {411}
  \mid r_n \, \,(t,\varepsilon) \mid \,\, \leq \, \, \frac{\varepsilon}{n}  \,\, \rho \,\,  e^{-a \,t} \ \ \ \  \ \ \ n \, \in [1, \,\, N]
\end{equation}

\vspace{3mm}
\noindent and so

\vspace{3mm}
\begin{equation}                \label {412}
  \rho^{-1} \, \,e^{at} \mid R_1 \mid \,\,\leq \, \, \frac{2  }{\pi} \,\,   \sum ^N_{n=1} \,\, \frac{\varepsilon }{n}\,\, \leq \,\, \frac{2\, \varepsilon}{\pi} \, ( c_0 \, +\ \frac{1}{2N}+ ln\,  N).
\end{equation}

\vspace{3mm} \noindent
For each positive 
 constant \@$\beta \, $, one has $\,\,\ ln \, N \, < \, \beta \,\, N ^{1/\beta}\,\,  $ so that  for $\, \beta \, = k^{-1} \,\,\,  (k<1)\,\, $ the estimate (\ref{47}) follows, with $\rho $ and $c_1$ defined by  (\ref{46}). As for $R_2$ one has:

\vspace{3mm}
\beq            \label{413}
  R_2  =   \sum ^\infty _{n=N+1} \frac{1}{n^2} \,\,  h_n ( t,\varepsilon )  g_n - e^{-at} \sum ^\infty _{n=N+1} \frac{sen(b_n t)}{n^2\, b_n}  g_n  = R'_2-R''_2
\eeq

\vspace{3mm}\noindent where the terms $h_n$ defined in (\ref{38}) \@ represent now hyperbolic functions $(B_n^2 <0)$. For this it results:

\vspace{3mm}
\beq                           \label{414}
 h_n(t,\varepsilon)\,\, \leq \,\, t \,\, \,\,\  e^{-\,\,\frac{c^2 n^2 }{\varepsilon n^2 +2a}\,\, t } \ \\ \  \\\,\,\,\,\, \ \ \forall n\geq N+1  
\eeq

\vspace{3mm}\noindent
and $ N+1 \geq 2c/\varepsilon $ \@ for \@ $ \varepsilon < 2(c-a). $ \@ As consequence:

\vspace{3mm}
\beq                             \label{415}
 \mid R'_2 \mid \,\, \leq \,\, \frac{2 \,t}{\pi} \,\,  e^{-\,\frac{c^2}{2} \, \frac{t}{\varepsilon}}\,  \sum ^\infty _{n=1} \,\, \frac{1}{n^2}\,\, =(1/3) \,\, \varepsilon \,\, \theta\,\, e^{-\,\frac{c^2}{2}\, \theta}, 
\eeq

 \vspace{3mm}\noindent 
with $\theta = t/\varepsilon$. At last, as $ b_n \,\, > b_1 \,n $, one deduces that

\vspace{3mm}
\beq                             \label{416}
 e^{at} \, \mid R''_2 \mid \,\, \leq \,\, \frac{2 }{\pi\, b_1} \,\, \sum ^\infty _{n=N+1} \,\, \frac{1}{n^3}\,\, \leq \frac{1}{3 b_1} \,\, (N+1)^{-1} \leq \,\, \frac{\pi \, \varepsilon }{3c \, b_1} . 
\eeq

 \vspace{3mm}\noindent 
and (\ref{415})-(\ref{416}) imply the estimate (\ref{48}).
\hbox{}\hspace*{5mm}
\rule{1.85mm}{1.85mm}

\vspace{3mm}
Referring to (\ref{48}), let observe that 

\vspace{3mm}
\beq                   \label{417}
\theta \, e^{-\frac{c^2}{2}\, \theta} \,\, \leq \,\,( 4/c^2  e
)\,\, 
e^{-\frac{c^2}{4}\,\theta}
\eeq

\vspace{3mm}\noindent
and let $ b\equiv \, min \, (a, c^2 /4 \varepsilon)$.Thus, by (\ref{46}),(\ref{47}),(\ref{48}) one has

\vspace{3mm}

\beq           \label{418}
\mid \, R_1 \, \mid \, +\, \mid \, R_2 \, \mid \,\, \leq \, \varepsilon ^k \, r(t) \, e^ {-b\,t}
\eeq

\vspace{3mm}\noindent
with

\vspace{3mm}
\beq                 \label {419}
r(t) \, =  \varepsilon ^{1-k} \,\, [\, c_1 (\varepsilon) \, \rho (t) \, + \, c_2 \, ( 1+ 4/c^2 e)].
\eeq

\vspace{3mm}\noindent
Then, the following theorem can be stated.

\vspace{3mm}
{\bf Theorem 4.1 }- {\em Whatever the positive constant }$ k<1$ {\em may be, for all}$\,\,t \,\geq \,0\,\,$ {\em and }$x \in [0,\pi]$, {\em it results}:

\vspace{3mm}
\beq                   \label{420}
\mid \, H_\varepsilon \, - \,H_0\, \mid \, \leq \, \gamma \, \varepsilon ^k \, ( 1+ t+ t^2) \, e^{-b\,t},
\eeq

\vspace{3mm}
\noindent
{\em where the constant } $\, \gamma $  {\em depends only on } $k,\,a,\,c$.

\vspace{3mm}
{\bf Remark 4.1}- The asymptotic analysis of this section and the results of theorem 4.1 can be applied also to the function $ K_\varepsilon $ defined by (\ref{318}) and related to the problem ${\cal H}_\varepsilon$.
 \vspace{5.1mm}

\section {Conclusions}
 \setcounter{equation}{0}

\hspace{5.1mm}

\vspace{3mm}

To outline  a first application and to avoid too many
 formulae, let consider only the term depending on the source. Then, referring to the problems   ${\,\cal
P}_\varepsilon \,$ and  ${\,\cal
P}_0\, $ and putting $\,\Delta H \, = H_\varepsilon \, -\, H_0\,$, it results

\vspace{3mm}

\begin{equation}                  \label {51}
u_{\varepsilon}\, -\, u_0\,= -\,  \int_{0}^{\pi} \, \,d \xi \,\,\int_{0}^{t} f_{\xi\xi} (\xi,\tau)\,\, \ {\Delta H} (x,\xi, t-\tau) \,d\tau.
\eeq

\vspace{3mm}
By assuming that $\, f  \in  C^2(\Omega)\, $ and that $\, f_{xx}\,$ is bounded also when $ \, t \, \rightarrow \, \infty\, $, let

\begin{equation}                  \label {52}
||\,u_f \,|| \, = \, \sup_\Omega \,\,|\,f_{xx} \, (x,t)\,|. 
\eeq

\vspace{3mm}\noindent
As consequence of theorem 4.1, when $\, \varepsilon \, \rightarrow 0\,$, the following rigorous approximation holds:

\vspace{3mm}

\begin{equation}                  \label{53}
u_\varepsilon \, =\, u_0\, + \, \varepsilon ^k \,\, r \,\,\,\\\ \ \ \ \\\ \ \ \  \forall (x,t) \in \Omega
\eeq  

\vspace{3mm}\noindent
where the error $\, r \, $ is such that

\vspace{3mm}

\begin{equation}                  \label{54}
|\, r\,| \, \leq \, \gamma _1 \, \,|| u _f|| \,\,\,\\\ \ \ \ \\\ \ \ \  \forall t \geq 0
\eeq  

\vspace{3mm}\noindent
and the constant $ \, \gamma _1 \, $ depends only by $ a,\,  \, c,\, k.\,$ So, {\em the error of the approximation is neglegible also for large } $ \, t \,\,(t \rightarrow \infty).$

\vspace{3mm}
When $\,f\,$   is non linear, an integral equation like 

\vspace{3mm}

\begin{equation}                  \label {51}
v \,= \int_{0}^{\pi}d \xi \int_{0}^{t}  {\Delta H} (x,\xi, t-\tau) F
[\xi,\tau, u(\xi, \tau ), u_\tau(\xi,\tau), u_\xi(\xi,\tau)]d\tau
\eeq

\vspace{3mm}\noindent
must be analyzed. These applications will be dealt successively.

\hspace{5.1mm}

\end{document}